\documentclass[referee,pdflatex,citeautoscript,sn-nature]{sn-jnl}

\usepackage[T1]{fontenc}
\usepackage[utf8]{inputenc}
\usepackage{graphicx}
\usepackage{amsmath,amssymb,bm}
\usepackage{hyperref}
\usepackage{booktabs}
\usepackage{siunitx}
\usepackage[section]{placeins} 

\newcommand{\TM}{\mathrm{TM}}

\newcommand{\AAng}{\text{\AA}}

\begin{document}

\title{Resonance-enhanced super--superexchange yields giant chiral magnon splitting in rutile altermagnets}

\author[1,2]{\fnm{Dai Q.} \sur{Ho}} 
\author[1]{\fnm{D.} \sur{Quang To}} 
\author[3]{\fnm{Byungkyun} \sur{Kang}} 
\author[1]{\fnm{Matthew F.} \sur{Doty}}
\author[4,5]{\fnm{Garnett W.} \sur{Bryant}} 
\author*[1]{\fnm{Anderson} \sur{Janotti}} 

\affil*[1]{\orgdiv{Department of Materials Science and Engineering},
\orgname{University of Delaware},
\orgaddress{\city{Newark}, \state{Delaware}, \postcode{19716}, \country{USA}}}

\affil[2]{\orgdiv{Faculty of Natural Sciences},
\orgname{Quy Nhon University},
\orgaddress{\city{Quy Nhon}, \postcode{55113}, \country{Vietnam}}}

\affil[3]{\orgdiv{Department of Physics},
\orgname{The University of Texas at El Paso},
\orgaddress{\city{El Paso}, \state{Texas}, \postcode{79968}, \country{USA}}}

\affil[4]{\orgdiv{Nanoscale Device Characterization Division, Joint Quantum Institute},
\orgname{National Institute of Standards and Technology},
\orgaddress{\city{Gaithersburg}, \state{Maryland}, \postcode{20899-8423}, \country{USA}}}

\affil[5]{\orgname{University of Maryland},
\orgaddress{\city{College Park}, \state{Maryland}, \postcode{20742}, \country{USA}}}

\abstract{Altermagnets host momentum-selective spin splitting and chiral-split magnonic excitations despite vanishing net magnetization, enabling spin-transport without ferromagnetism. In rutile structures, establishing altermagnetism spectroscopically has been challenging, motivating the search for a rutile platform with a resolvable exchange-driven chiral magnon splitting. Here we combine hybrid-functional first-principles calculations with linear spin-wave theory to show that rutile CuF\textsubscript{2} exhibits a meV-scale splitting between magnon modes of opposite chirality along momentum directions dictated by its $d$-wave altermagnetic symmetry. The splitting originates from an anomalously strong long-range super--superexchange channel Cu--F$\cdots$F--Cu, which enhances the symmetry-allowed difference between seventh-neighbour exchanges, $J_{7b}-J_{7a}$, controlling the chiral-mode splitting. We identify an orbital-resonance mechanism: energetic alignment between Cu $3d_{z^2}$ and F $2p_z$ states strengthens virtual hopping along the Cu--F$\cdots$F--Cu path and amplifies the anisotropic long-range exchange. Rutile CuF\textsubscript{2} therefore provides an ideal platform to validate rutile altermagnetism and suggests an orbital-energy description for engineering large chiral magnon splittings in insulating altermagnets.}


\maketitle


Altermagnetism is a magnetic phase characterized by large spin splitting in electronic and magnonic band structures without net magnetization, enabled by symmetry relations between magnetic sublattices that differ from those of conventional collinear antiferromagnets~\cite{vsmejkal2022beyond,vsmejkal2020crystal,vsmejkal2023chiral}.
Experimental signatures have been reported across multiple crystal classes, including MnTe~\cite{krempasky2024altermagnetic,lee2024broken,osumi2024observation,liu2024chiral}, MnTe$_2$~\cite{zhu2024observation}, and CrSb~\cite{reimers2024direct,yang2025three}, establishing altermagnetism as a broadly realized phase. Rutie is a particularly appealing structure, where theory predicts a momentum-dependent splitting and a characteristic $d$-wave pattern in their electronic structure and spin exciation (magnon) spectrum~\cite{vsmejkal2020crystal, yuan2020giant, vsmejkal2022beyond, vsmejkal2023chiral}.

Despite this promise, direct spectroscopic confirmation of rutile altermagnetism remains elusive. RuO$_2$ was proposed as a prototypical rutile altermagnet and predicted to enable an efficient nonrelativistic electrical spin-splitter response arising from momentum-selective spin splitting in the rutile motif~\cite{GonzalezHernandez2021SpinSplitter}, yet recent measurements suggest bulk RuO$_2$ is likely nonmagnetic and lacks direct evidence of intrinsic spin splitting~\cite{song2025absence,wu2025fermi}. Correspondingly, inelastic neutron scattering and related probes report either featureless spectra or broad continua rather than sharp chiral magnon branches~\cite{Kessler2024,kiefer2025crystal,yumnam2025constraints}. These results motivate the search for rutile materials with established antiferromagnetic order and an exchange-driven chirality splitting large enough to exceed dipolar interactions and instrumental resolution limits.

Transition-metal difluorides $\TM$F$_2$ ($\TM$= Mn, Fe, Co, Ni, Cu) crystallize in the rutile structure and are established antiferromagnets. Recent polarized-neutron studies report a minute chirality-dependent response in MnF$_2$~\cite{faure2025altermagnetism}, and closely related work discusses small magnon splittings in FeF$_2$~\cite{sears2026altermagnetic}. Here we show that rutile CuF$_2$ is exceptional: it exhibits a \emph{giant} (meV-scale) chiral-split magnon modes driven by a resonance-enhanced super--superexchange interaction.
We note that bulk CuF$_2$ is commonly reported in a monoclinic structure derived from the rutile parent, but the rutile phase remains a plausible metastable target that could be stabilized in thin films via epitaxial constraint on symmetry-matched substrates and/or kinetic control during growth~\cite{Aramburu2019InorgChem_CuF2_groundstate,Aramburu2019PCCP_CuF2_instability,Aykol2018ThermodynamicLimitMetastable}.



\begin{figure}
\centering
\includegraphics[width=0.75\linewidth]{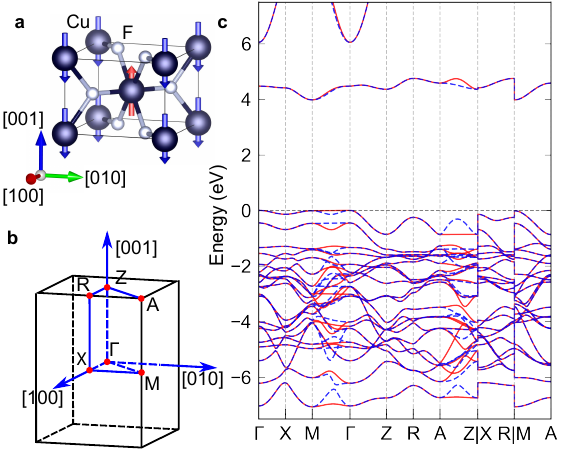}
\caption{\textbf{Electronic structure of rutile CuF$_2$.}
\textbf{a,} Rutile crystal structure and magnetic sublattices. \textbf{b,} Brillouin zone and high-symmetry paths. \textbf{c,} HSE06 band structure illustrating momentum-selective spin splitting characteristic of rutile altermagnets. The valence-band maximum is set to 0.}
\label{fig:fig1}
\end{figure}

Rutile $\TM$F$_2$ hosts two magnetic sublattices that are connected by a non-symmorphic $4_2$ screw rotation (rather than by inversion or any primitive lattice translation) combined with time-reversal operation, satisfying the symmetry conditions for altermagnetism~\cite{vsmejkal2022beyond}.
Figure~\ref{fig:fig1}a shows the rutile CuF$_2$ structure and its two magnetic sublattices. Hybrid-functional calculations (HSE06) yield a Mott-insulating electronic structure dominated near the gap by $\TM$ $3d$ and F $2p$ states (Fig.~\ref{fig:fig1}c), in good agreement with previous works on $\TM$F$_2$~\cite{correa2018electronic}. Importantly, consistent with rutile altermagnetic symmetry, bands are spin split along anisotropic directions such as $\Gamma$--M and Z--A while remaining spin degenerate along other high-symmetry lines.

\FloatBarrier


Because $\TM$F$_2$ are insulating with localized moments, their low-energy magnetism can be described by an extended Heisenberg model
\begin{equation}
\mathcal{H}=-\sum_{i\neq j}J_{ij}\,\mathbf{S}_i\cdot \mathbf{S}_j,
\label{eq:Heis}
\end{equation}
where long-range interactions are essential in rutile altermagnets~\cite{vsmejkal2023chiral}. We extract $J_{ij}$ using the Liechtenstein-Katsnelson-Antropov-Gubanov (LKAG) Green-function formalism~\cite{liechtenstein1987local} via the magnetic force theorem as implemented in \texttt{TB2J} package~\cite{he2021tb2j}. The resulting exchange parameters up to the seventh neighbours are summarized in Table~\ref{tab:J}.

\begin{table}[!t]
\caption{\textbf{Exchange parameters $J_i$ (meV) for rutile $\TM$F$_2$ extracted from first-principles calculations.}}
\centering
\begin{tabular}{lrrrrrrrr}
\toprule
$\TM$F$_2$ & $J_1$ & $J_2$ & $J_3$ & $J_4$ & $J_5$ & $J_6$ & $J_{7a}$ & $J_{7b}$\\
\midrule
MnF$_2$ & -0.503 & -2.317 & 0.003 & -0.006 & -0.001 & -0.006 &  0.003 & -0.023 \\
FeF$_2$ & -0.318 & -2.196 & 0.005 & -0.007 & -0.001 & -0.002 &  0.000 & -0.040 \\
CoF$_2$ & -0.349 & -1.324 & 0.004 & -0.004 & -0.001 & -0.005 &  0.002 & -0.030 \\
NiF$_2$ & -0.238 & -1.570 & 0.002 & -0.004 &  0.000 & -0.006 &  0.002 & -0.024 \\
CuF$_2$ & -0.318 & -0.806 & 0.025 & -0.014 &  0.000 & -0.002 & -0.009 & -0.232 \\
\bottomrule
\end{tabular}
\label{tab:J}
\end{table}
\FloatBarrier

\begin{figure}
\centering
\includegraphics[width=0.6\linewidth]{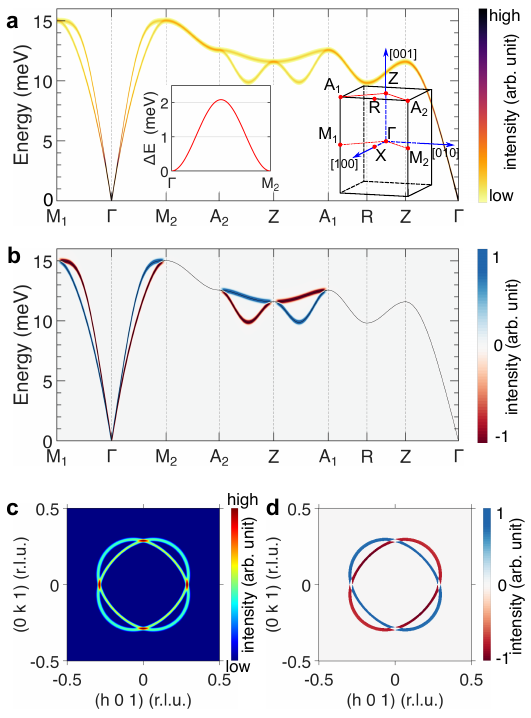}
\caption{\textbf{Chiral magnon spectrum of rutile CuF$_2$.}
\textbf{a,} Simulated inelastic neutron scattering intensity along high-symmetry paths (left inset: splitting magnitude; right inset: path definition). \textbf{b,} Spectrum weighted by neutron chiral factor, highlighting opposite chirality of the two modes. \textbf{c,} Constant-energy slice of the dynamic structure factor at 10 meV (r.l.u: reciprocal lattice unit). \textbf{d,} Corresponding chiral factor exhibiting the rutile $d$-wave pattern.}
\label{fig:fig2}
\end{figure}

We compute magnon dispersions and dynamic structure factors using linear spin-wave theory as implemented in \texttt{SpinW}~\cite{toth2015linear}. The simulated inelastic neutron scattering intensity for CuF$_2$ (Fig.~\ref{fig:fig2}a) shows two magnon branches that are nearly degenerate over most of the Brillouin zone but split strongly along rutile-anisotropic directions ($\Gamma$--M and Z--A). The neutron chiral factor reveals the opposite chirality of the two split modes and a $d$-wave pattern in constant-energy slices (Fig.~\ref{fig:fig2}b--d), consistent with altermagnetic properties in rutile structures.

A key feature of rutile altermagnets is that the chirality splitting of the two magnon modes requires symmetry-inequivalent long-range exchanges. In particular, the splitting is controlled by the difference between two seventh-neighbour paths, $J_{7b}-J_{7a}$~\cite{vsmejkal2023chiral,bandyopadhyay2025rational}.
The splitting was shown to follow the symmetry-derived form~\cite{bandyopadhyay2025rational}
\begin{equation}
\Delta \varepsilon^{\mathrm{magnon}}_{\mathbf{k}}
=\varepsilon_{\alpha}(\mathbf{k})-\varepsilon_{\beta}(\mathbf{k})
=4\,(J_{7b}-J_{7a})\,\sin(k_x a)\,\sin(k_y a),
\label{eq:splitting}
\end{equation}
For Mn--Ni, $J_{7a}$ and $J_{7b}$ are small and nearly cancel each other, leading to $\mu$eV-scale splittings. In contrast, CuF$_2$ exhibits a pronounced enhancement of $J_{7b}$, producing a much larger $J_{7b}-J_{7a}$ difference and therefore directly explains the meV-scale chiral magnon splitting as seen in Fig.~\ref{fig:fig2}a.
This splitting magnitude is comparable to that recently observed experimentally in $g$-wave altermagnets MnTe~\cite{liu2024chiral} and $\alpha$-Fe$_2$O$_3$~\cite{hoyer2025altermagnetic}, and is large enough to unambiguously identify anisotropic exchange coupling as the dominant mechanism---as opposed to the competing long-range magnetic dipole-dipole interaction known to be present in this material series~\cite{faure2025altermagnetism, sears2026altermagnetic}.  

\FloatBarrier

\begin{figure}
\centering
\includegraphics[width=0.8\linewidth]{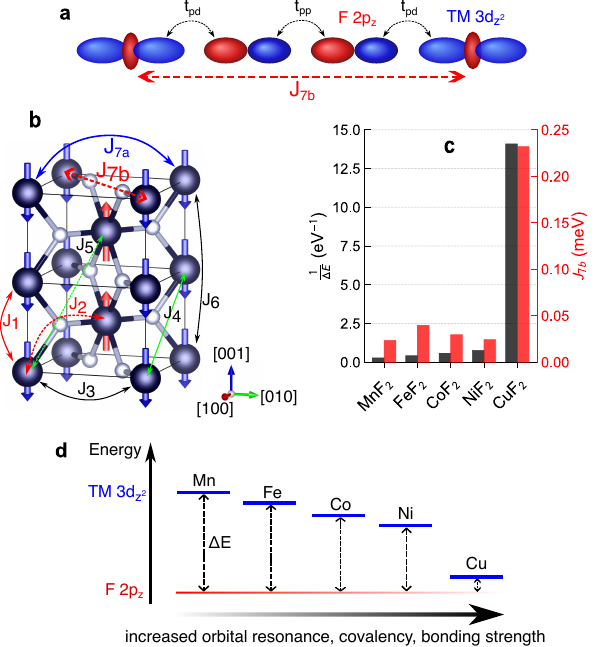}
\caption{\textbf{Origin of enhanced $J_{7b}$ in rutile CuF$_2$.}
\textbf{a,} Orbitals contributing to the dominant super--superexchange channel for $J_{7b}$ and \textbf{b,} Spin-exchange paths up to seventh neighbours.  \textbf{c,} Correlation between $J_{7b}$ and $1/\Delta E$ with $\Delta E=\varepsilon(\TM\,3d_{z^2})-\varepsilon(\mathrm{F}\,2p_z)$. \textbf{d,} Schematic orbital-energy alignment highlighting enhanced resonance for CuF$_2$.}
\label{fig:fig3}
\end{figure}

$J_{7b}$ is anomalously large in CuF$_2$ despite its long Cu--Cu separation along the seventh-neighbour path. The relevant exchange channel is a linear super--superexchange pathway $\TM$--F$\cdots$F--$\TM$ along $\langle 110\rangle$ (Fig.~\ref{fig:fig3}a,b). In this geometry, virtual hopping is dominated by $\sigma$-type overlap between $\TM$ $3d_{z^2}$ and F $2p_z$ states, supplemented by F--F $p$--$p$ hopping. Perturbatively, the effective exchange amplitude is enhanced as the orbital-energy offset $|\varepsilon_p-\varepsilon_d|$ decreases (details in Supplementary Information Section~S2; classic background in Refs.~\cite{anderson1950antiferromagnetism,Anderson1959}).

Consistent with this picture, the $\TM$F$_2$ series exhibits increasing energetic overlap between $\TM$ $3d$ and F $2p$ states from Mn to Cu. Figure~\ref{fig:fig3}c correlates $J_{7b}$ with the inverse energy offset $\Delta E^{-1}$, where $\Delta E=\varepsilon(\TM\,3d_{z^2})-\varepsilon(\mathrm{F}\,2p_z)$ obtained from Wannier onsite energies. CuF$_2$ exhibits the strongest orbital alignment (resonance), strengthening the anisotropic long-range exchange responsible for the giant chiral magnon splitting.

\FloatBarrier

\begin{figure}
\centering
\includegraphics[width=0.65\linewidth]{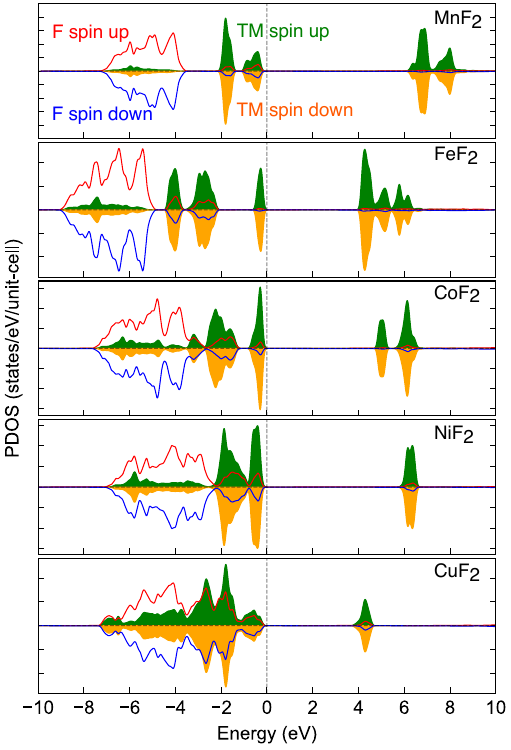}
\caption{\textbf{Projected density of states across rutile $\TM$F$_2$.}
Atomic- and spin-resolved PDOS comparing $\TM$ $3d$ and F $2p$ states. The energetic overlap (orbital resonance) increases from Mn to Cu and is strongest for CuF$_2$, consistent with enhanced covalency and strengthened super--superexchange. The valence-band maximum is set to 0 for each compound.}
\label{fig:fig4}
\end{figure}

Furthermore, the projected density of states (PDOS) (Fig.~\ref{fig:fig4}) reveals a systematic increase in energetic overlap between the $\TM$ $3d$ and F $2p$ manifolds from MnF$_2$ to CuF$_2$, indicating progressively stronger $\TM$--F covalency and supporting a resonance-enhanced super--superexchange picture. In MnF$_2$ the Mn $3d$-dominated states are largely separated from the F $2p$-dominated valence bands, whereas for CuF$_2$ the Cu $3d$ and F $2p$ characters become strongly intermingled across the valence region. The comparable Cu and F contributions over the same energy window imply substantial Cu--F hybridization, which increases the matrix elements for the long-range pathway $\TM$--F$\cdots$F--$\TM$ and thereby amplifies the anisotropic super--superexchange. This electronic-structure trend is consistent with the chemical-bonding analysis in Supplementary Information (Section~S3), which shows enhanced covalency for Cu--F bonds through larger integrated projected crystal order hamiltonian populations (\(-\)IpCOHP), higher bond order through the integrated crystal order bond index (ICOBI), and a reduced formal charge state.

\FloatBarrier

Our results establish rutile CuF\textsubscript{2} as a particularly transparent insulating platform in which the hallmark symmetry of rutile altermagnetism produces a large, momentum-selective splitting between opposite-chirality magnon modes. This design principle is complementary to recent proposals for ferroic control of altermagnetism, including ferroelastic switching of altermagnetic spin splitting in CuF$_2$ and related systems~\cite{peng2025ferroelastic}. By linking the splitting to the symmetry-allowed anisotropic exchange difference \((J_{7b}-J_{7a})\), we identify a concrete microscopic lever---a long-range $\TM$--F$\cdots$F--$\TM$ super--superexchange channel whose strength is dramatically increased by orbital resonance between $\TM$ $3d$ and ligand $2p$ states. The same resonance criterion provides a practical descriptor for materials design: aligning the relevant $\TM$ $3d$ level (here Cu $3d_{z^2}$) with the ligand $2p$ manifold enhances covalency, strengthens the long-range pathway, and amplifies anisotropic exchanges and chiral magnon splittings.

\textit{Experimental outlook.} The predicted chiral-mode splitting is largest along the anisotropic directions of the rutile structure where the exchange anisotropy is most pronounced, and it yields two nondegenerate magnon branches with distinct neutron chiral factors. A direct test is therefore high-resolution inelastic neutron scattering on single crystals, targeting the momentum cuts where the splitting is maximal, i.e., in the middle of the $[hhl]$ momentum paths, and comparing the measured dispersion and spectral weight to the calculated neutron response. Polarization analysis (when available) provides an additional handle to isolate the chiral contribution to the magnetic scattering and to map the characteristic rutile ``$d$-wave'' chirality pattern. Complementary probes such as RIXS (at the $\TM$ edge) or THz spectroscopy can access the same low-energy spin-wave modes in regimes where sample volume or momentum coverage is limiting. Together, these measurements would provide a stringent, model-resolved validation of rutile altermagnetic magnons and, more broadly, of resonance-enhanced super--superexchange as a route to large anisotropic exchanges in insulating antiferromagnets.


\section*{Methods}

\subsection*{First-principles calculations}
Bulk $\TM$F$_2$ was modeled in the tetragonal rutile structure ($P4_2/mnm$, space group 136).
The lattice parameters were optimized using first-principles calculations (listed in the Supplementary Information), showing good agreement with experimental data~\cite{wyckoff1963crystal, baur1971rutile}.
Spin-polarized Density Functional Theory calculations used the HSE06 screened hybrid functional~\cite{heyd2003hybrid,krukau2006influence} as implemented in VASP~\cite{ref25vasp1996CMS,ref26vasp1996PRB}.
Projector augmented-wave (PAW) datasets were employed to treat the ion-core and valence shell interaction~\cite{ref27_Blochl1994PAW,ref28kresse1999ultrasoft}. The plane-wave cutoff energy was set to 600~eV. Brillouin-zone sampling used $\Gamma$-centered meshes with \texttt{KSPACING}=0.25~\AAng$^{-1}$ for structural optimization and a denser mesh with \texttt{KSPACING}=0.20~\AAng$^{-1}$ for electronic structure calculations. Electronic convergence was set to $10^{-6}$~eV.

\subsection*{Wannierization and exchange parameters extraction}
Electronic structures obtained from VASP calculations were downfolded to a Wannier basis Hamiltonian spanning the $\TM$ $3d$ and F $2p$ manifold (22 Wannier orbitals per unit cell). Wannier orbital onsite energies were extracted and used to define $\Delta E$ in Fig.~\ref{fig:fig3}c. Exchange parameters were computed using the LKAG Green-function method~\cite{liechtenstein1987local} based on the magnetic force theorem as implemented in \texttt{TB2J}~\cite{he2021tb2j}. 

\subsection*{Spin-wave calculations and neutron response}
Magnon spectra and dynamical structure factors were computed using linear spin-wave theory as implemented in \texttt{SpinW}~\cite{toth2015linear}. Simulated neutron intensities were convoluted with a Gaussian broadening of FWHM 0.2~meV.

\section*{Data availability}
The data supporting the findings of this study are available from the corresponding authors upon reasonable request.

\section*{Code availability}
Calculations used \texttt{VASP}, \texttt{Wannier90}, \texttt{TB2J}, and \texttt{SpinW}. Scripts used for post-processing are available from the corresponding authors upon reasonable request.

\section*{Acknowledgements}
This work was supported by the NSF through the UD-CHARM University of Delaware Materials Research Science and Engineering Center (MRSEC) Grant No.~DMR-2011824.  We also acknowledge the use of computational resources from the National Energy Research Scientific Computing Center (NERSC), a Department of Energy Office of Science User Facility, through the NERSC award BES-ERCAP 0034471 (m5002), and the DARWIN computing system at the University of Delaware, which is supported by the NSF Grant No.~1919839.

\section*{Author contributions}
D.Q.H and D.Q.T contributed equally to this work. D.Q.H. performed first-principles calculations and magnon modelling (Wannierization, spin-wave, and neutron-related calculations). D.Q.T. assisted in formulating the problem and in conducting analytical analysis for exchange interactions. B.K.K. M.F.D., G.W.B., and A.J. contributed to interpretation, discussion, and supervision of the project. All authors wrote and revised the manuscript.

\section*{Competing interests}
The authors declare no competing interests.


\end{document}